# Magnetic crossover and complex excitation spectrum of the ferromagnetic/antiferromagnetic spin-1/2 chain system α-TeVO$_4$

V. Gnezdilov,[1] P. Lemmens,†[2] A. A. Zvyagin,[1,3] V. O. Cheranovskii,[4] K. Lamonova,[5] Yu. G. Pashkevich,[5] R. K. Kremer,[6] and H. Berger[7]

[1] B.I. Verkin Inst. for Low Temperature Physics and Enginering, NASU, 61103 Kharkov, Ukraine;

[2] Inst. For Condensed Matter Physics, TU Braunschweig, D-38106 Braunschweig, Germany;

[3] Max-Planck Institute for Physics of Complex Systems, D-01187 Dresden, Germany;

Max-Planck-Institut für Festkörperforschung, 70569 Stuttgar, Germany;

[4] V.N. Karazin Kharkov National University, 61077 Kharkov, Ukraine;

[5] A.A. Galkin Donetsk Phystech NASU, 83114 Donetsk, Ukraine;

[6] Max-Planck-Institut für Festkörperforschung, 70569 Stuttgart, Germany;

[7] Inst. Phys. Mat. Complexe, EPFL, CH-1015 Lausanne, Switzerland

We report magnetic susceptibility, specific heat, and Raman scattering investigations of α-TeVO$_4$ containing V-O edge-sharing chains. These chains promote a system of ferromagnetic/antiferromagnetic spin-1/2 Heisenberg alternating exchange chains with pronounced spin frustration. The magnetic susceptibility and Raman scattering evidence a crossover at $T^* = 85$ K with different slopes of the reciprocal susceptibility and a magnetic phase transition into a long-range-ordered state at $T_c = 16$ K. From Raman scattering data a strong mutual coupling between lattice and magnetic degrees of freedom is deduced. A comparison to model calculations and prior Raman scattering on other chain systems yield a plausible interpretation of the microscopic mechanism for the crossover behavior.

# I. INTRODUCTION

One-dimensional (1D) quantum spin systems have been of interest and investigated both experimentally and theoretically for many years. Studies of chain systems are important for testing various theoretical concepts and approximations. They exhibit a rich phase diagram and unconventional magnetic properties originating from low dimensionality and pronounced quantum fluctuations.

In real compounds low dimensional magnetic behavior is usually a consequence of large differences between the magnitudes and signs of the exchange couplings between neighboring magnetic ions located in different directions. Much is known about the ground-state properties, the magnetic phase diagram, and the excitation spectrum of quantum spin-1/2 chains with both nearest- (NN) the next-nearest-neighbor (NNN) antiferromagnetic (AFM) interactions ($J_1 > 0$, $J_2 > 0$).[1-6] Relatively less is known about low dimensional systems with ferromagnetic (FM) NN and antiferromagnetic NNN interactions ($J_1 < 0$, $J_2 > 0$). Though the corresponding model has been a subject of some studies[7-11] the complete picture of the phases of this model is unclear. Recently, $LiCuVO_4$ [12], $Rb_2Cu_2Mo_3O_{12}$ [13] and $Li_2ZrCuO_4$ [14] with edge-sharing chains of copper-oxide elementary units have been discovered as representing 1D spin-1/2 Heisenberg system with ferromagnetic NN and antiferromagnetic NNN competing exchange interactions. For $LiCuVO_4$ the exchange parameters have been determined by inelastic neutron scattering as $J_1 = -19$ K and $J_2 = 65$ K with $\alpha = J_2/J_1 \approx -3$ and long range AFM ordering has been observed for temperatures below about 2.5 K. For $Rb_2Cu_2Mo_3O_{12}$ the exchange parameters were estimated as $J_1 = -138$ K and $J_2 = 51$ K with $\alpha = J_2/J_1 \approx -0.37$. No long range magnetic order has been discovered down to 2 K so far, so the compound has been considered as a model system. The compound $Li_2ZrCuO_4$ has a ratio $\alpha \approx -0.3$ which is close to the critical value $\alpha_c = -1/4$ and shows evidence for magnetic ordering at $T_m = 6.4$ K.[14] These and related studies reactivated the theoretical interest in low-dimensional systems with the FM NN and AFM NNN interactions [14-17] especially as ferromagnetic NN interaction is expected to exist in a wide class of transitional metal compounds with edge-sharing $MeO_4$ units.

While the ground state properties vs. $J_2/J_1$ ratio in the FM-AFM case have been discussed in scientific literature, the excitation spectrum is not yet studied and understood completely. Using Raman spectroscopy technique, we have studied the magnetic excitation spectrum in frustrated alternating ferromagnetic quantum spin-1/2 chain system of $\alpha$-TeVO$_4$. Though this compound is known for a long time[18] to our present knowledge, its physical properties at all were hitherto not reported. Here, we present an investigation of the magnetic exciations studied by Raman spectroscopy. In addition, the temperature dependencies of magnetic susceptibility $\chi(T)$ and specific heat $C_p(T)$ were measured and analyzed, too.

## II. EXPERIMENTAL DETAILS

The growth of $\alpha$-TeVO$_4$ single crystals is described elsewhere.[18] Samples with dimension 3×3×8 mm$^3$ were oriented by x-ray Laue diffractometry. Raman scattering measurements were performed in quasi-backscattering geometry with the excitation line $\lambda$ = 514.5 nm of an Ar$^+$ laser. The laser power of 10 mW was focused to a 0.1 mm diameter spot on the sample surface. Spectra of the scattered radiation were collected by a DILOR-XY triple spectrometer and recorded by a nitrogen cooled charge-coupled device detector with a spectral resolution of <0.5 cm$^{-1}$. Magnetic susceptibility, $\chi(T)$, of $\alpha$-TeVO$_4$ was measured in a range $1.85 \leq T \leq 330$ K for 0.1 T and 5 T by a SQUID magnetometer (Quantum Design). Specific heat measurements were performed using a PPMS calorimeter (Quantum Design) using the relaxation method.

## III. RESULTS AND DISCUSSION

### A. Structure and important parameters

The compound $\alpha$-TeVO$_4$ crystallizes in a monoclinic structure (P2$_1$/c) with the lattice parameters $a$ = 5.099 Å, $b$ = 4.93 Å, $c$ = 12.672 Å, and $\beta$ = 105.85 Å with Z = 4 formula units TeVO$_4$ per unit cell at room temperature.[18] The crystal structure of $\alpha$-TeVO$_4$ is shown in Figure 1(a). The structure consists of $[VO_4]_n^{4n-}$ zigzag chains parallel to the $b$ axis formed by distorted VO$_6$ octahedra sharing edges. The lone pair cation Te$^{4+}$ leads to a magnetic separation of chains with respect to each other. Figure 1(b) shows the topology of V$^{4+}$ and O$^{2-}$ ions in the crystal structure of $\alpha$-TeVO$_4$ forming chains with nonlinear exchange path. It is clearly seen that the NN V-V bond in the chains has an alternation: a V-V distance is 2.87 Å and V-O-V angle $\theta$ = 97.07° in one bond; another distance is 3.27 Å with the angle $\theta$ = 107.29°. We assume that the NN exchange interactions in these bonds, $J_1$ and $J_1'$, respectively, have different magnitudes due to the difference in V-V distances and V-O-V angles.

Mizuno et al.[19] analyzed the magnetic interactions angle dependencies by diagonalizing a three band Hubbard Hamiltonian for finite copper oxide clusters with edge shared oxygen atoms. They came to the conclusion that the angle at which FM exchange interaction achieves its maximum is close to 87° while the critical angle where the exchange interaction changes its sign is close to 95°. In a more elaborate *ab-initio* approach developed by de Graaf et al.[20] it was found that the critical angle is close to 104° while FM nearest-neighbor exchange reaches a maximum at 97°.

According to de Graaf et al.[20] and our room temperature structural data the sign of $J_1$($J_1'$) is presumed to be FM(AFM) at higher temperatures, although one cannot ignore the fact that the value of 107.29° is close to the critical angle of ~104° at which the nearest-neighbor interaction is expected to change sign.[20] In addition to $J_1$($J_1'$), NNN super-superexchange interaction $J_2$, which originates from the V-O-O-V path, also play an important role in the magnetic properties. The interaction $J_2$ for the edge-sharing case is generally AFM (>0) and its magnitude is known to be of a few ten Kelvin.[16,19] Thus, the structural data allow us to consider $\alpha$-TeVO$_4$ as a spin-1/2 chain system possibly with competing (spin frustrating) NN and NNN interactions and an alternating NN exchange interaction.

### B. Magnetic susceptibility and specific heat

The magnetic susceptibility $\chi(T)$ of $\alpha$-TeVO$_4$ measured along and perpendicular to the $b$ crystallographic direction is shown in Figure 2(a). The temperature dependence of $\chi(T)$ shows a sharp maximum near 17 K. A Curie-Weiss fit of the high-temperature susceptibility for 110 K < $T$ < 300 K yields a Curie constant $C$ = 0.344 emu·Kmol$^{-1}$ and a positive Curie-Weiss temperature of $\Theta_{CW}$ = +25.6 K for magnetic fields applied along the $b$ direction. For magnetic fields applied in the $ac$ plane we find C = 0.346 emu·Kmol$^{-1}$ and $\Theta_{CW}$ = +24.8 K. Positive Curie-Weiss temperatures indicate predominant ferromagnetic interactions. At $T_c$ = 16 K, the kink in $\chi(T)$ may indicate that the compound undergoes a magnetic phase transition into a long-range-ordered phase. The kink in $\chi(T)$ becomes more evident in a plot of $d\chi/dT$ vs. $T$ as a sharp peak shown in Figure 2(b). The temperature of the maximum in $\chi(T)$, $T_{max}^{\chi}$ = 18.4 K, has been taken as the temperature at which $d\chi/dT = 0$. The presence of both NN and NNN interactions (implied by the structure, for the analysis see below) affects the dependence of the ordering temperature on the inter-chain and intra-chain interactions.[21] It is yet impossible, unfortunately, to find the correct expression for the ordering temperature for the proposed quasi-one-dimensional model, consisting of weakly coupled spin chains with NN and NNN interactions.

Figure 2(c) presents a plot of 1/$\chi(T)$ vs. $T$ in a magnetic field of 0.1 T along the chains direction. The solid line is a Curie-Weiss law which fits very well to the data down to $T \approx$ 110 K. At lower temperatures, 1/$\chi(T)$ shows a gradual (~40 K wide) crossover centered at $T^* \approx$ 85 K with a different slope above and below this temperature region. Note that the dotted line, namely the linear fit of 1/$\chi(T)$ at temperatures below the crossover region, intercept the abscissa axis at 4.6 K. The same picture is valid for $\chi(T)$ with a magnetic field perpendicular to chains direction.

Below we will try to relate the crossover behavior analyzing structural and electronic properties of the compound.

Taking into account the magnetic susceptibility data and a closer look on the lattice structure of α-TeVO$_4$ we will try to determine the sign of $J_1$ and $J_1'$. The structure analysis suggests that the sign of $J_1$ is presumably FM, while $J_1'$ is probably AFM.[20] It is necessary to point out, that since the nearest neighbor V-V bonds have two alternating configurations, additional antisymmetric exchange interactions, such as Dzyaloshinskii-Moriya or magnetically anisotropic interactions, together with an alternating g-tensor, may be important in this material, too. A similar assumption was made during the analysis of the thermodynamic properties of the edge-sharing copper oxide Rb$_2$Cu$_2$Mo$_3$O$_{12}$.[13,16] However, as it will be shown below, a FM $J_1$ and AFM $J_1'$ do not agree with the temperature behavior of the magnetic susceptibility, cf. Figure 2, in the studied compound.

The zero-field specific heat $C_p(T)$ of a α-TeVO$_4$ single crystal (between 2 K and 83 K) is shown in Figure 3. The insert shows $C_p/T$ vs. $T$ with a λ-shaped peak with the maximum at the temperature $T_{\max}^{C_p}$ = 16.9 K. This characteristic temperature nearly coincides with the maximum temperature in $d\chi/dT$. Besides that, the observed ratio of $T_{\max}^{\chi}/T_{\max}^{C_p}$ = 1.09 differs from the predicted (1.33)[22] for AFM chains and the experimentally observed in FM-AFM chains of Li$_2$ZrCuO$_4$ (1.17)[17] and the quasi-1D FM Cu-peptides (1.73, 1.84).[23] Both facts indicate that the sharp peak in $C_p(T)$ can be attributed to a magnetic phase transition. An inspection of the λ-shaped peak in the inset in Figure 3 reveals that it contains an entropy of 2.5 J/molK or ~40% of Rln2, with R being the molar gas constant, corresponding to the entropy of a $S = 1/2$ system. Apparently, ~60% of the entropy is removed in short-range correlations above 16.9 K. In order to estimate the exchange parameter modeled to the heat capacity with the following approach: We assume that above ~20 K the magnetic contribution to the heat capacity can be approximated by a sum of the heat capacity of an AFM Heisenberg chain with uniform nearest-neighbor exchange coupling $J_{NN}$ according to Eqs. (54a) and (54b) in Ref. 2, and of a phonon contribution. The phonon part is represented by an extended Debye model. Accordingly, the total heat capacity was fitted to

$$C_p(T) = C_{\text{mag}}(J_{NN}) + C_{\text{ph}} = C_{\text{mag}}(J_{NN}) + T\sum_{n=1}^{4} a_n T^{2n} \quad (T \geq 20 \text{ K}),$$

with the coefficients $a_i$ and the exchange constant $J_{NN}$ as fit parameters. The best fits indicating $J_{NN}$ ~80 K is displayed by the solid lines in Figure 3 as the magnetic and the sum of magnetic and phonon contributions, respectively. The short range-order contributions contribute essentially around 40 – 60 K where the slope of $1/\chi$ differs from these one at $T > 110$ K. About

2/3 of the entropy are removed above 20 K in a built-up of short range ordering leaving about 1/3 to be removed in the long-range ordering, in agreement with the experimental finding. On the other hand, as it will be shown below, the temperature behavior of the magnetic susceptibility of the AFM spin-1/2 chain does not agree with the one, observed in the experiment, see Figure 2.

At low temperatures (2 K ≤ $T$ ≤ 5 K) below the λ-peak we observe a $C_p \propto T^{2.65}$ law. The exponent 2.65 being in between a 3D AFM (3) and a quasi-2D AFM (2) state, suggests that the transition at $T_c \approx 16$ K can be attributed to a transition into a long-range-ordered phase with a nontrivial magnetic structure. A comparison of our experimental results with those found for other zigzag chain structures [24-26] allows us to assume that at low temperatures $\alpha$-TeVO$_4$ has an incommensurate helimagnetic ground state.

The temperature dependences of the magnetic susceptibility and the magnetic specific heat cannot be fitted by known expressions for the homogeneous Heisenberg spin-1/2 chain assuming only NN AFM interactions, cf. Figure 4, where the results of our calculations are presented for several studied spin chain models. To analyze the temperature behavior of spin chains we used an exact diagonalization for small clusters of spins (short spin chains). In our calculations we used from 8 to 14 spins (even numbers) in the chains. This number is limited by the exponential growth of computation time. The accuracy of our calculations was sufficient to reproduce features of the temperature dependence of the specific heat and the magnetic susceptibility of the studied models down to temperatures of order of 0.01-0.1 (depending on the model) of the value of the NN exchange constant. We used arbitrary units; the values of the effective $g$-factor of the V$^{4+}$ ion, Bohr's magneton and Boltzmann's constants are taken to be equal to 1. From the red circles in Figure 4 one can see that: (i) the magnetic susceptibility of the homogeneous AFM chain manifests a maximum at higher temperatures, than the observed one, and, more importantly, (ii) the inverse susceptibility shows an AFM-like behavior (i.e. the effective Curie temperature is *negative*), unlike the observed experimental features, cf. Figure 2(c). Notice that the spin-gap-like behavior for $T < 0.05$ is due to finite-size effects,[22] while for infinite spin-1/2 AFM Heisenberg chains the magnetic susceptibility is finite and the specific heat is proportional to $T$ at low temperatures.[1] If one introduces an alternation of the NN interactions (as the structure of the material suggests) in the AFM chain it results in the onset of a spin gap in the low-energy excitations.[2] Alternating FM-AFM NN exchange interaction also leads to a singlet ground state with gapped excitations. The gap implies an exponentially small low-temperature magnetic susceptibility and specific heat, which does not agree with the observed features, cf. Figures 2 and 3. On the other hand, the alternation of FM NN exchange constants produces a divergent behavior of the susceptibility at low temperatures (the ferromagnetic ground state). Hence, alternating NN interactions alone cannot explain the features of the behavior of the studied system. Our next step is to introduce weak (as suggested by the structure) NNN AFM interactions between spins together with FM interactions between NN spins. We choose a FM sign of the NN interactions, to reproduce the FM-like behavior of the inverse susceptibility. Figure 4 presents the temperature behavior of the magnetic susceptibility, inverse susceptibility and the specific heat for $J_1 = -1$, $J_1' = -0.9$, and $J_2 = 0.1$ (wine red, right-handed triangles). It is clear that the behavior is different from the one, observed in the experimental data: While the FM character of the high-temperature magnetic susceptibility is present (in accordance with the observed features, cf. Figure 2(c)), the low-temperature behavior also manifests a FM behavior, i.e. the divergence of $\chi(T)$, which is not the case in the studied system. Notice the two-maxima structure of the temperature dependence of the specific heat, which is characteristic for spin-1/2 chains with FM NN and AFM NNN interactions.[17] For larger values of $J_2 = 0.3$ the low-temperature magnetic susceptibility is small, but the high-temperature inverse magnetic susceptibility shows a negative Curie temperature, which also contradicts the experimentally observed features, see Figure 4 (green stars). Figure 4 (violet up-directed triangles) shows the temperature behavior of thermodynamic characteristics of the model with alternating NN interactions both in values and in sign, namely, the FM $J_1 = -1$, the AFM $J_1' = 0.9$, and with the NNN AFM interaction, $J_2 = 0.1$. One can see that again, the observed behavior (cf. Figs. 2,3) does not agree with the one of the characteristics of this model – the model clearly shows features, characteristic for spin-gapped systems. Finally, black squares present the temperature characteristics of a model with alternating in sign and magnitude NN interactions with $J_2 = 0.3$ ($J_1 > -4 J_2$),[17] and this model also manifests a typical spin-gap behavior. It also turns out that our exact diagonalization results for short spin-1/2 chains with alternating FM NN interactions and AFM NNN interactions agree with recent analytical calculations.[27]

This is why we suppose that the Hamiltonian of a single spin-1/2 chain of $V^{4+}$ ions has the form:

$$H = \sum_{i=1}^{N} \left[ J_1^z S_{2i}^z S_{2i-1}^z + J_1^{xy} \left( S_{2i}^x S_{2i-1}^x + S_{2i}^y S_{2i-1}^y \right) \right] + $$
$$+ \sum_{i=1}^{N} \left[ J_1'^z S_{2i}^z S_{2i+1}^z + J_1'^{xy} \left( S_{2i}^x S_{2i+1}^x + S_{2i}^y S_{2i+1}^y \right) \right] + $$
$$+ \sum_{i=1}^{N} \left[ J_2^z S_{2i-1}^z S_{2i+1}^z + J_2^{xy} \left( S_{2i-1}^x S_{2i+1}^x + S_{2i-1}^y S_{2i+1}^y \right) \right] + $$
$$+ \sum_{i=1}^{N} \left[ J_2^z S_{2i}^z S_{2i+2}^z + J_2^{xy} \left( S_{2i}^x S_{2i+2}^x + S_{2i}^y S_{2i+2}^y \right) \right] \quad . \quad (1)$$

Here $J_1^z$, $J_1^{xy}$, $J_1'^z$, and $J_1'^{xy}$ are the alternating exchange constants between the nearest-neighbor spins 1/2, $J_2^z$ and $J_2^{xy}$ are the exchange constants of the interaction between next to nearest neighboring spins, and $S_i$ denote operators of spin-1/2 at the $i$-th site of the chain. Because of the alternation of the lattice spacing between nearest $V^{+4}$ ions the model takes into account the possible (small) alternation of the NN exchange couplings. We also introduce a small uniaxial magnetic anisotropy of the exchange interactions $J^z \neq J^{xy}$, which follows from, e.g., the different temperature dependence of the magnetic susceptibility of the system along and perpendicular to the $b$ axis, see Figure 2(a). Notice that the mentioned temperature behavior implies an "easy-plane" type of the magnetic anisotropy. Figure 4 (blue, down-directed triangles) presents the temperature behavior of the magnetic susceptibility, the inverse magnetic susceptibility, and the magnetic specific heat for the considered model with the parameter values $J_1^z = -1$, $J_1^{xy} = -1.09$, $J_1'^z = -0.9$, $J_1'^{xy} = -0.99$, $J_2^z = 0.1$, and $J_2^{xy} = 0.109$. The calculations were performed for a chain with up to 14 spins. Figure 4(d) shows how the temperature dependence of the inverse magnetic susceptibility depends on the size of the studied chain model. On can see, that the difference in the behavior of spin chains of lengths 10, 12, and 14 is very small, and it is revealed only for low temperatures. Summarizing, our choice of parameters for the spin-1/2 chain model Hamiltonian is based on the following features of the temperature behavior of the real system. First, our model reproduces the AFM-like finite value of the magnetic susceptibility at low temperatures. Second, the FM-like behavior of the high-temperature part of the magnetic susceptibility is also reproduced by the model: The inverse susceptibility at high temperatures can be fitted to a Curie-Weiss law with a positive (i.e., ferromagnetic) Curie temperature. Also, similar to what was observed in the experiment, the inverse susceptibility manifests a gradual crossover to much weaker FM-like behavior (or AFM-like behavior) at intermediate temperatures. Finally, the temperature dependence of the specific heat of the model reveals two maxima (as expected, cf. Ref. 17). The low-temperature maximum is sharp, and it takes place approximately at the same temperature, at which the magnetic susceptibility shows a maximum, $T_{\max}^\chi / T_{\max}^{C_p} \approx 1$, as the experiment implies. The second maximum of the specific heat is smoother than the first one, and it is situated at intermediate temperatures, approximately at which the crossover in the inverse magnetic susceptibility $\chi^{-1}(T)$ takes place. We have to point out that our choice of parameters is suggested by the crystal structure, and it is the minimal possible choice to reproduce the observed experimental features, because taking only an alternation of the NN interactions and/or NNN interactions into account cannot provide a qualitative agreement with the experimental data. We emphasize that the situation with low-temperature maxima of the magnetic susceptibility and the specific heat caused by one-dimensional interactions between spins are close to the temperature of a phase transition to a magnetically ordered state was already discussed for a quasi-one-dimensional spin-1/2 system with FM NN and AFM NNN couplings.[14] It turns also out that because of the presence of the first maximum in the temperature dependence of the magnetic specific heat the standard mean-field feature of the possible transition into a magnetically ordered phase can be affected by that maximum, and, also, the common AFM three-dimensional mean-field exponent (3) can be affected by the one-dimensional one (1), leading to a reduction of the observed value of the exponent in real materials. The behavior of the magnetic susceptibility and the specific heat of the model with Hamiltonian (1) implies, that it has (i) a singlet ground state, and (ii) two possible branches of excitations, with the lowest one being gapless. Excitations, belonging to this lowest branch can be called spinons because their properties are similar to spinon excitations of the homogeneous AFM Heisenberg chain, see below.

Such a behavior of the model may be related to the existence of a quantum critical point. Because of the frustrating FM-NN and AFM-NNN spin-spin interactions it is expected that there is a quantum phase transition dividing the incommensurate and commensurate phases.[17] According to a study of an integrable spin-1/2 chain with NN and NNN spin-frustrating interactions,[28] the quantum critical point can produce shifts of the maxima of the magnetic susceptibility and the specific heat of spin-1/2 chain with NN and NNN interactions to lower temperatures compared to standard models with only NN spin-spin couplings, which agrees with the results of our exact diagonalization for short spin chains.

Although the initial $T$-dependence of the magnetic susceptibility qualitatively agrees with the experimental one, the sign of the resolved NN intrachain coupling $J_1'$ may disagree with the angle based estimations of Refs. 19,20 for copper oxide spin chain compounds. Also, we stress that our analysis yields only a qualitative, but not quantitative agreement with the experimental data (for short chains the positions of maxima in the temperature behavior of the magnetic susceptibility and specific heat can be shifted, compared with the ones for long chains). We also attribute these deviations to simplifications of the model as longer-ranged[24-26] and some intrachain interactions are possibly lacking.

### C. Phonons

Polarized Raman spectra of $\alpha$-TeVO$_4$ measured at temperatures 290 K and 5 K are shown in Figure 5. The sharpness of the observed phonon modes indicates a high quality of our single crystal. The monoclinic (P2$_1$/c, Z = 4) crystal structure of $\alpha$-TeVO$_4$ with all atoms having a site symmetry of 4e leads to $\Gamma = 18A_g(aa, bb, cc, ac) + 18B_g(ab, bc) + 17A_u + 16B_u$ Raman- and infrared-active phonon modes.

Experimentally, in the frequency region of 10 – 1400 cm$^{-1}$, thirty two phonon modes were identified in the spectra and their behavior was analyzed. In Figure 6 the result of a temperature analysis of representative phonons is shown. With decreasing temperature several distinctive features show up. First, upon cooling from room temperature all modes undergo hardening and then a saturation in frequency for temperatures around $T^*$ (Fig. 6(a)). Upon further cooling they show a jump down at $T = 50$ K and further hardening at lower temperatures. Second, the linewidths of the phonons show an anomalous behavior at temperatures below $T^*$ especially for the phonon lines with lower frequency (Fig. 6(b)). Normally, phonon linewidths narrow monotonously with decreasing temperatures. Third, the integrated intensity of phonon lines shows an anomalous behavior (Fig. 6(c)). Summarizing, the characteristic temperatures where anomalies in $\chi(T)$ and $C_p(T)$ are seen are also evident in the phonon spectra. This suggests a significant spin-phonon coupling in $\alpha$-TeVO$_4$.

### D. Magnetic Raman scattering

Magnetic scattering in α-TeVO$_4$ is evident as quasielastic scattering and as distinct, finite energy modes, both with characteristic temperature dependencies in intensity and frequency. Firstly, we will discuss the quasi-elastic Raman response as shown in Figure 7. Our experimental setup was suitably adjusted so that Rayleigh scattering is suppressed for frequencies above $\omega > 12$ cm$^{-1}$ and the observed scattering is therefore intrinsic. Such a quasielastic scattering contribution may be due to spin diffusion[29,30] or fluctuations of the energy density of the spin system.[31] The former mechanism leads to a Gaussian lineshape[32] of the central line while the latter to a Lorentzian[31] and it is important for systems with non-negligible spin-phonon coupling, as given for α-TeVO$_4$ due to the anomalies shown in Figure 6. Spin-phonon coupling leads to an enhancement of the spectral weight of the energy fluctuations by reducing their time scale.[33] In addition, the Lorentzian spectral function is in very good agreement with our observed quasielastic linewidth.

Due to the action of the scattering Hamiltonian on a 1D spin system this scattering contribution should only be observed in intrachain scattering configuration, i.e. with the incident and scattered light polarization parallel to the chain direction. Experimentally this is not the case for α-TeVO$_4$ as this signal is observed with even stronger intensity in crossed polarizations. We attribute this violation of the selection rules to the orientation of the nearest neighbor V-V bonds and V-O-V-O planes (see Fig. 1(b)) that alternate. The chains are distorted into a zigzag shape with an effective direction along the $b$ axis. This chain geometry, namely a deflection of V-V exchange pathes from the $b$ direction, leads to a violation of the light scattering selection rule as it is governed by local hopping processes. We note a similar infringement for the spin chain compound (VO$_2$)P$_2$O$_7$ [34] which is attributed to two dimensional correlations including an additional diagonal AF exchange.[35]

For the intrachain *(bb)* and interchain *(aa)* scattering configurations the quasielastic Raman response decreases smoothly with lowering temperatures (see Fig. 7(a)). In the crossed polarizations *(ab)* there is an abrupt decrease of the scattering intensity at the crossover temperature $T^*$ (see Fig. 7(b)) and a possible further decrease for $T < T^*$. We attribute these effects to sudden changes of the energy density fluctuations as discussed further below.

The phonon lines in Raman spectra of α-TeVO$_4$ at $T > T_c$ are superposed by a structurized, temperature and symmetry dependent background (see Fig. 5). The large width of the observed signal distinguishes it from the comparably sharp phonon lines. Raman active transitions between crystal split *d* levels of the V$^{4+}$ ions should have a larger energy.[36,37] We have subtracted phonon lines leading to Figure 8. It is evident that the signal remains broad with decreasing temperatures and does not reflect the discrete nature of excitations between well defined atomic electronic levels. We therefore assume magnetic excitations and the corresponding two-magnon Raman scattering processes as its origin similar to other chain systems.[38,39] In the following we will discuss the polarization and temperature dependence of such scattering.

In our model (1) the ground state is given by a spin-singlet and it has low-energy gapless spinon-like excitations. Spinons carry a spin of 1/2 and their dynamical structure factor is given by a gapless two-particle continuum restricted by a lower and an upper dispersing boundary.[40,41] Light scattering leads to total spin-zero excitations with total momentum $k = 0$, e.g. two- or four-spin excitations. As the spectral weight for two spinons at $k = 0$ vanishes the excitation spectrum consists of four-spin excitations with $k = 0$ and an energy range up to $\omega = 2\pi J$.

Alternating or frustrating the coupling induces a quantum phase transition from a gapless critical into a gapped spin liquid state in the magnetically isotropic model. For $J_2 > 0$, the spin gap opens with $J_2/J_1 > 0.241$.[42-44] When $J_1 < 0$ and $J_2 > 0$ with $-0.25 < J_2/J_1 \leq 0$, the ground state is fully ferromagnetic, and becomes a singlet incommensurate state for $J_2/J_1 < -0.25$.[7] It is suggested that in this incommensurate state the gap is strongly suppressed.[45] These features are generic for spin-1/2 models with spin frustration NN and NNN interactions (see also Ref. 41), which low-energy excitations are gapless, and we can speculate that the model (1) also reveals the behavior of correlations, similar to what was studied in Ref. 46. Numerical calculations of the Raman intensity corresponding to the four spinon excitations for the one-dimensional spin-1/2 model in the parameter range of $-0.5 \leq J_2/J_1 \leq 0.5$ reveal a broad continuum like feature (Fig. 1 in Ref. 46). A comparison of these results with our Raman spectra in *(aa)* and *(bb)* scattering geometries leads to a qualitative agreement supporting the intuitive attribution of the $T > T_c$ continuum to spinon scattering. In a strictly 1D system, the Fleury-London polarization selection rules[47] do not allow coupling a perpendicular electronic polarization, e.g. the *(aa)* scattering configuration, to the chain direction. The observed scattering in (*aa*) and (*bb*) geometries is therefore attributed to the bent exchange path with contributions both parallel and perpendicular to the crystallographic *b* axis direction. In (*ab*) crossed polarization we observe only a featureless high temperature Raman band at around 100 cm$^{-1}$. This effect we attribute to the different form factor and the much weaker two-dimensional correlations that contribute to this scattering polarization As an example we refer to the polarization dependence of the magnetic scattering in the 2D cuprates.[48,49]

In the following we discuss the temperature dependence of the magnetic scattering, i.e. its evolution from broad continua to sharper modes at $T < T_c$ and anomalies for $T_c < T < T^*$ (see Fig. 8). We remark a few effects regarding these spectra. The first is a shift with cooling of the Raman band in the 100 cm$^{-1}$ region to higher frequencies. The second effect is the shift of a band in the 300 cm$^{-1}$ region to higher frequencies and its broadening. And the third effect is the temperature independence in position and width of the band in the region of 450 cm$^{-1}$. Theoretical modeling (see Fig. 4 in Ref. 46) shows that with decreasing temperatures the spectral features in the magnetic continuum shift to higher frequencies. This is roughly consistent with our experimental observations. It should be noted that in the 2D Heisenberg model the two-magnon peak broadens massively with increasing temperature with only a small reduction in frequency.[50]

The temperature dependence shown in Figure 8 correlates also with the characteristic temperatures obtained from magnetic susceptibility and specific heat measurements. In particular, the peak position of the band at 100 cm$^{-1}$ hardens linearly with cooling with an abrupt change of the slope at $T^* = 85$ K (see inset in Fig. 8). Beside that, its bandwidth changes noticeably with crossing $T^*$: It decreases in (*bb*) and (*aa*) geometries and increases in (*ab*) geometry. These observations together with the modeling of magnetic Raman continua as function of $J_2/J_1$ (Fig. 1 in Ref. 46) allow us to suggest that the crossover temperature $T^*$ is related to a modification of the exchange interaction along the zigzag chain. This process is also related to the phonon anomalies observed at $T^*$ and indicates a relevance of magneto-elastic coupling. Nevertheless a long range structural distortions for $T < T^*$ is not supported by the data as no new phonon modes are observed. On the other hand all ions are located on a general type 4e-position with identity being the only symmetry operation, i.e. shifts of the ions in the primitive cell do not violate space symmetry. The chain geometry implies that shifts that modify the critical V-O-V bond angle would lead to a change in the slope of $\chi^{-1}(T)$. Possible displacements are an enlargement of the O1-O1' distance (2.4 Å above $T^*$) which is the shortest oxygen-oxygen distance and shifts of the vanadium ions which reduce the alternation of short and long V-V intra-chain distances. These distortions are depicted in Figure 1. In spite of the similar action of O1 and V1 shifting on the V-O-V bond angles they lead to different changes of components of g-factor. Therefore we propose ESR studies to distinguish combined O1 and V1 shifts that should reduce the $g_{yy}$ and $g_{xx}$ components from O1 only shifts hat leaves the $g_{xx}$ component almost unchanged.

Figures 7 and 8 show that the Raman spectra demonstrate drastic changes also at low frequencies and temperatures below 16 K: (i) two sharp peaks appear around 37 and 47 cm$^{-1}$ and (ii) a peak appears at 62 cm$^{-1}$ with a linewidth of ~45 cm$^{-1}$. The latter feature is present in parallel and crossed polarizations with different peak intensity. The temperature dependence of the signals is analyzed in detail in Figure 9. The higher frequency mode is renormalized but persists well into the paramagnetic state. A very similar observation has been made in the helically ordered spin-chain systems LiCu$_2$O$_2$ and NaCu$_2$O$_2$ and interpreted as two magnon scattering and damping of short-range spin correlations by thermal fluctuations.[51,52] The low energy peaks at 37 and 47 cm$^{-1}$ are present only below the magnetic ordering temperature suggesting a one-magnon excitation as their origin. Based on their different temperature dependence we attribute them to acoustic and optical transverse magnons at $q = 0$. For a further analysis including the higher energy modes at 175 cm$^{-1}$ and the double-peak feature extending from ~300 cm$^{-1}$ to ~550 cm$^{-1}$ neutron scattering or Raman scattering under external magnetic field would be helpful. The complexity of this magnetic excitation spectrum is based on the helical spin correlations and the four-atom basis of the magnetic unit cell present in $α$-TeVO$_4$. A symmetry analysis of possible magnetic states in the possible ordered phase of α-TeVO$_4$ is presented in the Appendix.

## IV.    CONCLUSIONS

To summarize, we have presented magnetic susceptibility, specific heat, and Raman scattering data of the quasi-one-dimensional spin-1/2 chain system $α$-TeVO$_4$ with alternating NN interactions, and next-nearest-neighbor interaction. A Curie-Weiss fit of the magnetic susceptibility for $T > 80$ K yields a positive Curie-Weiss temperature $Θ_{CW} = +25.6$ K, indicating predominant ferromagnetic interactions. At $T^* \approx 85$ K the inverse magnetic susceptibility shows a crossover indicating a modification of the exchange interactions. The low-temperature magnetic susceptibility is finite, indicating antiferromagnetic correlations. Hence, the studied compound is a system in which ferro- and antiferromagnetic interactions compete. The observed features in the behavior of the specific heat, the magnetic susceptibility, and Raman scattering at $T_c = 16$ K can be interpreted as a phase transition of the studied quasi-one-dimensional spin system to a long-range ordered magnetic state.

A fit of the magnetic susceptibility and the specific heat in terms of single spin-1/2 chain model was performed. The best qualitative agreement with the experiment was obtained for the

FM alternating NN couplings and AFM NNN couplings with a weak easy-plane magnetic anisotropy. However, one has to keep in mind that the knowledge of $\chi(T)$ and $C_p(T)$ are not sufficient to determine all model parameters.[53] Usually, quasi-one-dimensional quantum spin systems with gapless low-energy excitations of their one-dimensional subsystems order if weak inter-chain interactions exist, see, e.g. References [54-56]. However, as it was shown in Reference [21], competing intra-chain spin-spin couplings drastically change the expressions for the ordering temperature and produce incommensurate magnetically ordered structures. Our case, however, is not covered by these expressions even if we have determined the characteristic parameters of the one-dimensional spin subsystem. This is due to the interplay of competing exchange interactions with spin-phonon coupling. Nevertheless, further investigations, especially inelastic neutron scattering, would be helpful to determine the exact magnetic structure of the ground state and the coupling parameters in α-TeVO$_4$ for temperatures above and below $T^*$. We highlight that the feasibility of large single crystal growth of α-TeVO$_4$ enables such and other studies.

Phonon Raman scattering indicate strong spin-lattice coupling by revealing distinct anomalies at $T^*$ and $T_c$. The unusually rich magnetic Raman spectrum of α-TeVO$_4$ was analyzed in a large temperature interval. The origin of these modes was discussed. Summarizing, we conclude that quantum spin systems with FM-AFM competing interactions in one and two dimensions shows very interesting phenomena. This is due to one part to the proximity to quantum critical points and by the other part to the nontrivial interplay of spin and lattice degrees of freedom. We have demonstrated in our study that for α-TeVO$_4$ this interplay is essential to understand its magnetic behavior.

ACKNOWLEDGMENTS

This work was supported by the DFG and the ESF network *Highly Frustrated Magnetism*. V.G. acknowledges the support of Ukrainian Program "Nanostructural systems, nanomaterials, and nanotechnology" and Ukr.-Rus. grant 2008-8; V.O.C. and A.A.Z. acknowledge the support from the Ukrainian Fundamental Research State Fund (F25.4/13).

APPENDIX

To perform a symmetry analysis of possible magnetic ordered states in α-TeVO$_4$ we follow the approach of Bertaut[54] and Izyumov and Naish.[55] The primitive cell of the α-TeVO$_4$

(space group P2$_1$/c) contains four V$^{4+}$ ions on 4e positions and coordinations are shown in Figure 1.

We introduce magnetic modes as linear combinations of sublattice spins S$_\alpha$, where $\alpha$ denotes a particular sublattice:

$$F = S_1+S_2+S_3+S_4 = m_1+m_2$$

$$L_1 = S_1+S_2 - S_3 - S_4 = m_1-m_2$$

$$L_2 = S_1-S_2 +S_3 -S_4 = l_1 + l_2$$

$$L_3 = S_1 -S_2 -S_3 +S_4 = l_1 - l_2$$

Here $\vec{l}_1 = \vec{S}_1 - \vec{S}_2$ and $\vec{l}_2 = \vec{S}_3 - \vec{S}_4$ denote AFM vectors and $\vec{m}_1 = \vec{S}_1 + \vec{S}_2$ and $\vec{m}_2 = \vec{S}_3 + \vec{S}_4$ the sublattice magnetizations of neighboring chains. **F** is the "ferromagnetism vector" of the crystal. **L$_1$** is determined by the difference of the ferromagnetism vectors of neighboring chains. **L$_2$** and **L$_3$** represent intrachain AFM ordering. For a second order magnetic phase transition the possible magnetic structures can be classified by the irreducible representations of the symmetry group of the crystal in the paramagnetic phase. The results of the symmetry operations are summarized in Table 1 where the first column contains irreducible representations of the P2$_1$/c space group. The corresponding symmetry operations and the basis vectors of magnetic structure are listed in the second and the third column, respectively. The last column presents the permutation symmetry of the magnetic modes. For an uniform magnetic order in a monoclinic Heisenberg magnet only one basis vector describes the magnetic structure in exchange approximation since the leading isotropic exchange is much stronger than Dzyaloshinskii-Moriya (DM) and anisotropic interactions. In the case of the four sublattice magnet the average magnitude of such a vector in the ordered state will be close to 4|S| while the other ones belonging to the same irreducible representation will be smaller by order of D/J.

Table 1. Symmetry of magnetic modes in α-TeVO$_4$

| C$_{2h}$(2/m) | 1 | 2$_y$ | I | m$_y$ | | |
|---|---|---|---|---|---|---|
| A$_g$ | 1 | 1 | 1 | 1 | L$_{1x}$, F$_y$, L$_{1z}$ | **F** |
| A$_u$ | 1 | 1 | -1 | -1 | L$_{2x}$, L$_{3y}$, L$_{2z}$ | **L$_3$** |
| B$_g$ | 1 | -1 | 1 | -1 | F$_x$, L$_{1y}$, F$_z$ | **L$_1$** |
| B$_u$ | 1 | -1 | -1 | 1 | L$_{3x}$, L$_{2y}$, L$_{3z}$ | **L$_2$** |

As it is apparent from Table 1, that components of the **L$_2$** and **L$_3$** vectors and the ferromagnetism vector **F** do not coexist in the same irreducible representation. This implies that weak ferromagnetism is incompatible with uniform AFM ordering in the chains. A detailed investigation of the exchange paths in the chain network shows that intra-chain DM interaction is absent for the given chain. For instance, both V(1)-O(1)-V(2) and V(1)-O(1')-V(2) exchange paths are symmetric and belong to the same plane, therefore the DM vectors of every paths have opposite directions and compensate each other. In the case of **L₁** type of Neel state with AFM order of nonzero ferromagnetic moments **m** on neighboring chains the weak ferromagnetism should appear only due to interchain interactions. Note that the zigzag-like geometry makes interchain interaction along *c* axis strongly asymmetric and frustrated.

One can show that any kind of uniform antiferromagnetic order will be unstable against the creation of incommensurate spin density waves. This is described by Lifshitz invariants which are allowed in this compound even in exchange approximation. In the Ginzburg-Landau approach they have the form:

$$\vec{m}_1 \frac{d\vec{l}_1}{dy} - \vec{m}_2 \frac{d\vec{l}_2}{dy}; \quad \vec{m}_1 \frac{d\vec{l}_1}{dz,x} + \vec{m}_2 \frac{d\vec{l}_2}{dz,x};$$

where the first term describes two anti-phase conical helixes on neighboring chains with helix vectors along *b*-axis. The microscopic origin of these invariants results from competing interactions and the frustration.

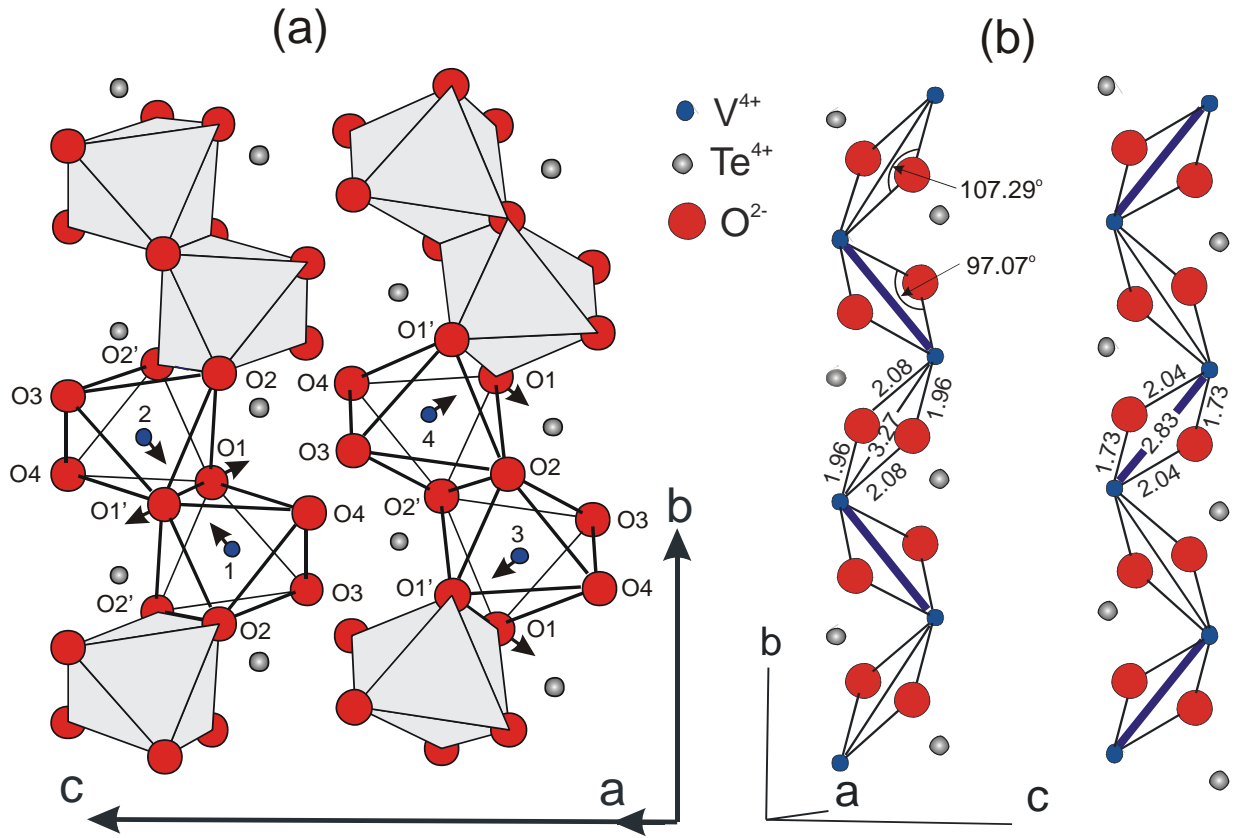

Figure 1. (Color online) (a) Projection of the lattice structure of α-TeVO$_4$. Chains along the *b* axis are formed by rows of VO$_6$ octahedra sharing edges. The vanadium ions are labeled from 1 to 4. We choose those site positions with the following coordinates (1) (0.380(1), 0.223(1), 0.434(4)); (2) (0.620(1), 0.777(1), 0.566(4)); (3) (0.380(1), 0.277(1), -0.066(4)); (4) (0.620(1), 0.723(1), 0.066(4)). Arrows show possible lattice distortions which change the angles of V-O-V bonds at *T* close to *T** = 85 K. (b) Schematic drawing of V$^{4+}$- and O$^{2-}$-ion positions in α-TeVO$_4$.

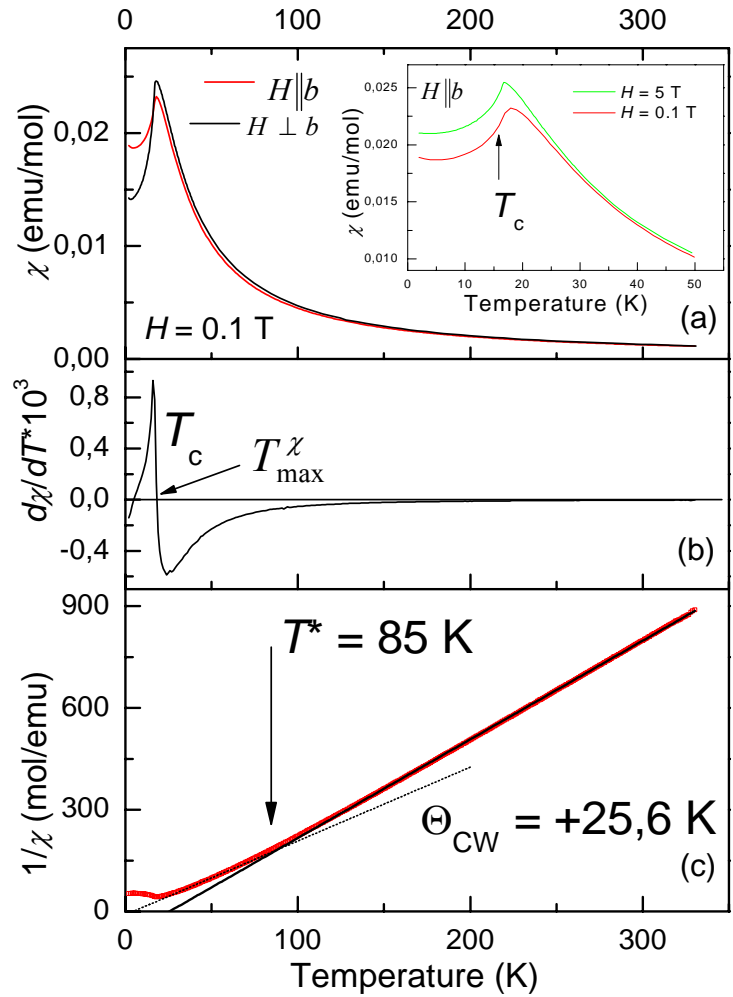

Figure 2. (Color online) (a) Temperature dependence of the magnetic susceptibility of α-TeVO$_4$ measured in a magnetic field $H = 0.1$ T applied along and perpendicular to the $b$ crystallographic axis. The inset shows susceptibility below 50 K measured in different magnetic fields. (b) $d\chi/dT$ versus $T$; the magnetic phase transition is manifested as a sharp peak at $T = 16$ K in $d\chi/dT$. (c) $1/\chi(T)$ versus $T$; the lines are the result of a Curie-Weiss fit.

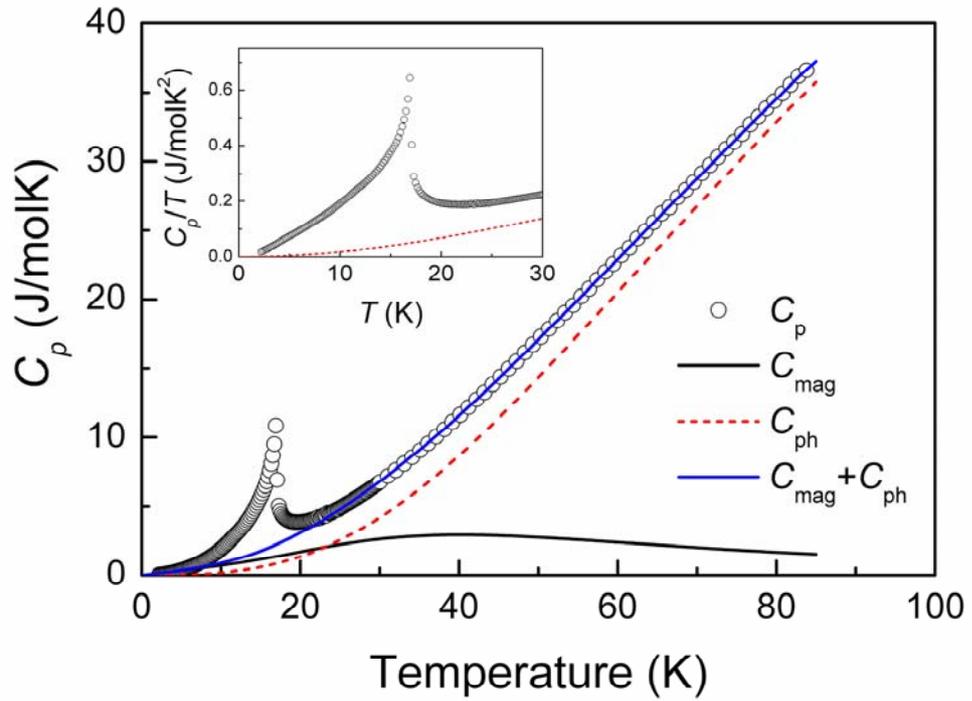

Figure 3. (Color online) The temperature dependence of the zero-field specific heat of α-TeVO$_4$ (open squares); the dotted line indicates the estimated phonon contribution $C_{ph}$. The solid lines give the magnetic contribution $C_{mag}$ of short range fluctuations and the sum of magnetic $C_{mag}$ and phononic $C_{ph}$ contributions, respectively. The inset shows $C_p/T$ vs. $T$ at low temperatures.

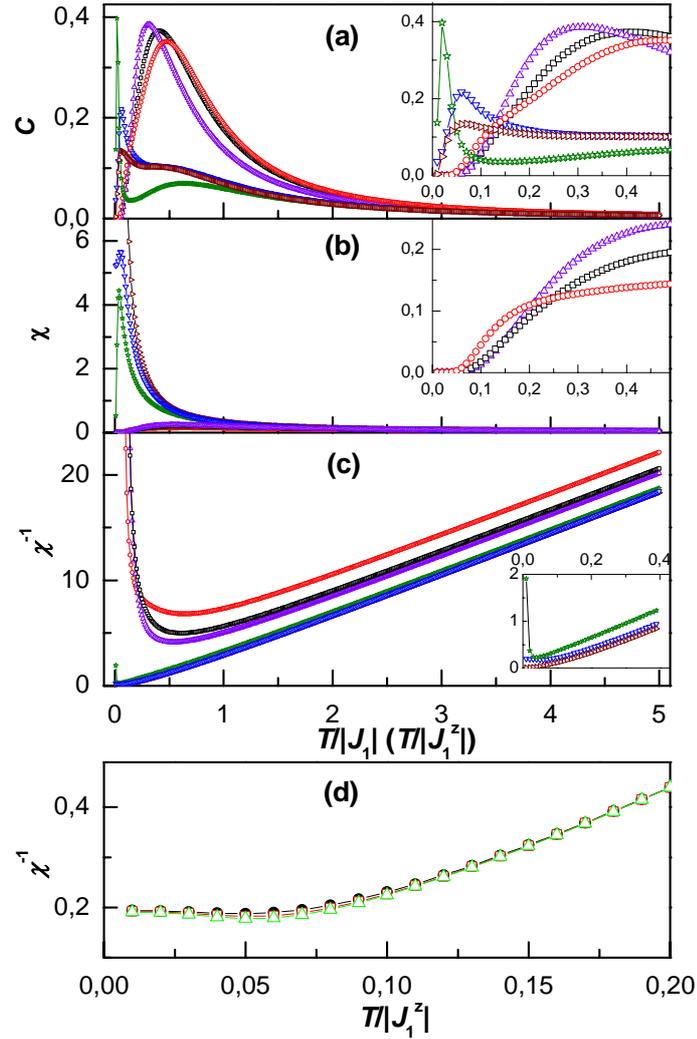

Fig. 4. (Color online) Calculated temperature dependencies of (a) the specific heat $C_p^{\text{magn}}$, (b) the magnetic susceptibility $\chi$, (c) the inverse magnetic susceptibility $1/\chi$, and for the several spin ½ chain models: the Heisenberg model with isotropic AFM NN interactions (red circles); a model with isotropic FM alternating NN interactions and weak AFM NNN interactions (wine red, right-handed triangles); a similar model with stronger AFM NNN interactions (green stars); a model with alternating in sign and magnitude NN interactions and weak AFM NNN ones (violet up-directed triangles); a similar model with stronger AFM NNN interactions (black squares); a model with magnetically anisotropic FM alternating NN interactions and weak AFM NNN ones (blue down-directed triangles), for details see text. Insets show the low-temperature evolution. Figure (d) shows the low-temperature dependence of the inverse magnetic susceptibility of the magnetically anisotropic model with FM alternating NN couplings and weak AFM NNN ones with spins ½ and chains with 10 (black circles), 12 (red squares) and 14 (green triangles) spins, respectively.

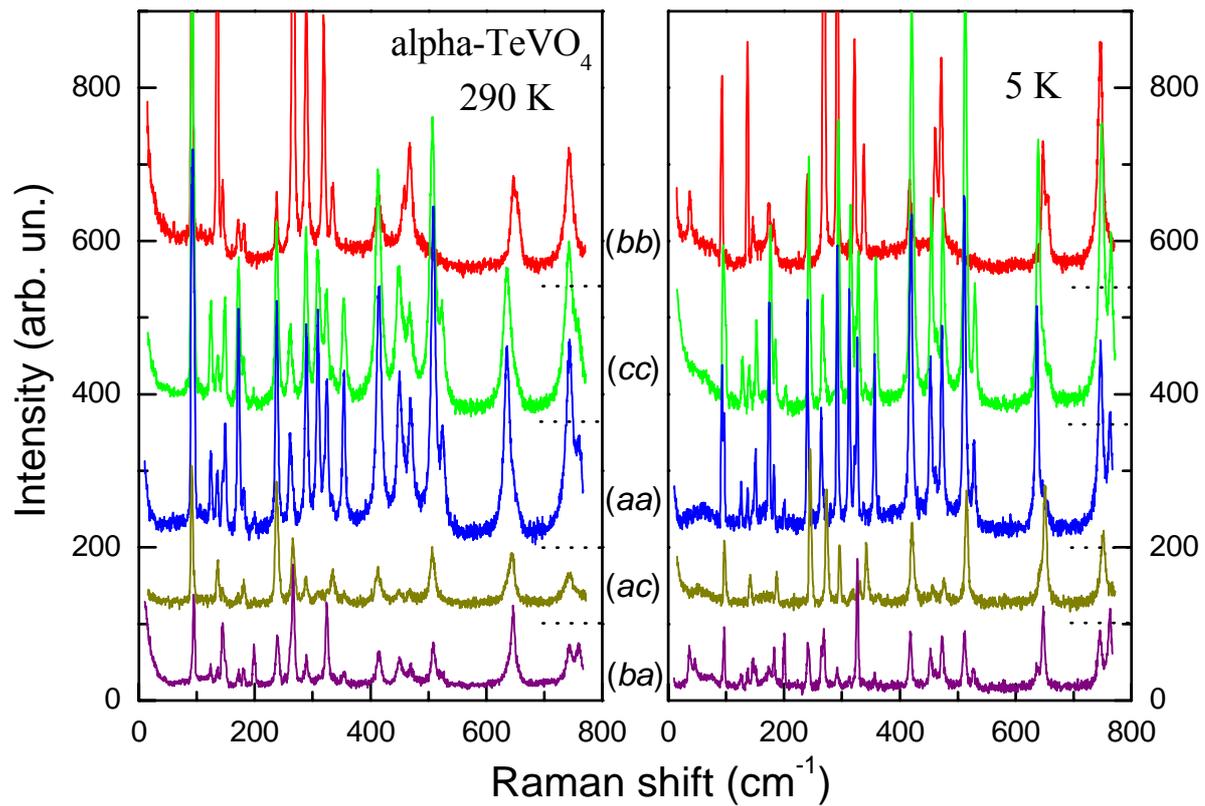

Figure 5. (Color online) Raman spectra of α-TeVO4 in different polarizations taken at 290 K and 5 K. Dotted lines represent the baselines of the vertically shifted data.

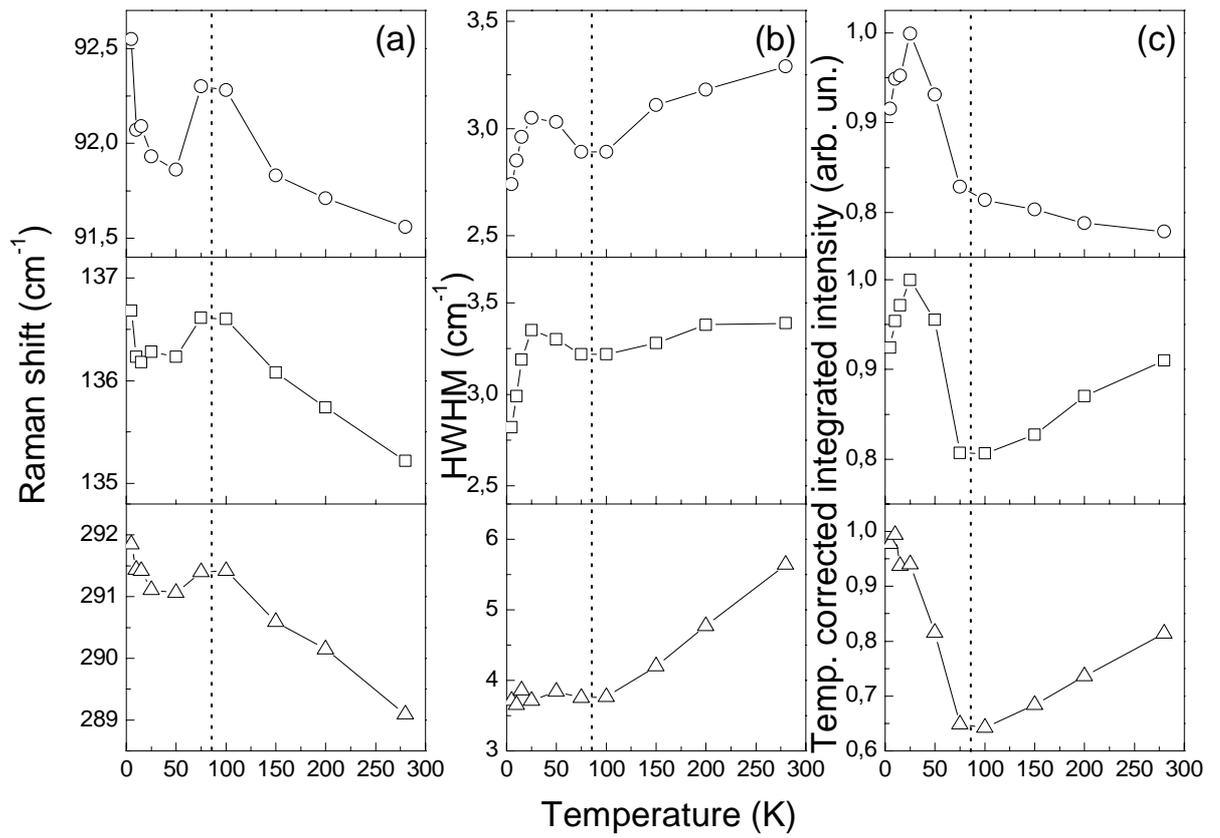

Figure 6. Temperature dependence of the peak frequencies (a), linewidths (b), and temperature corrected integrated intensities (c) of selected phonon lines in (*bb*) scattering configuration. Dotted lines indicate the crossover temperature $T^*$.

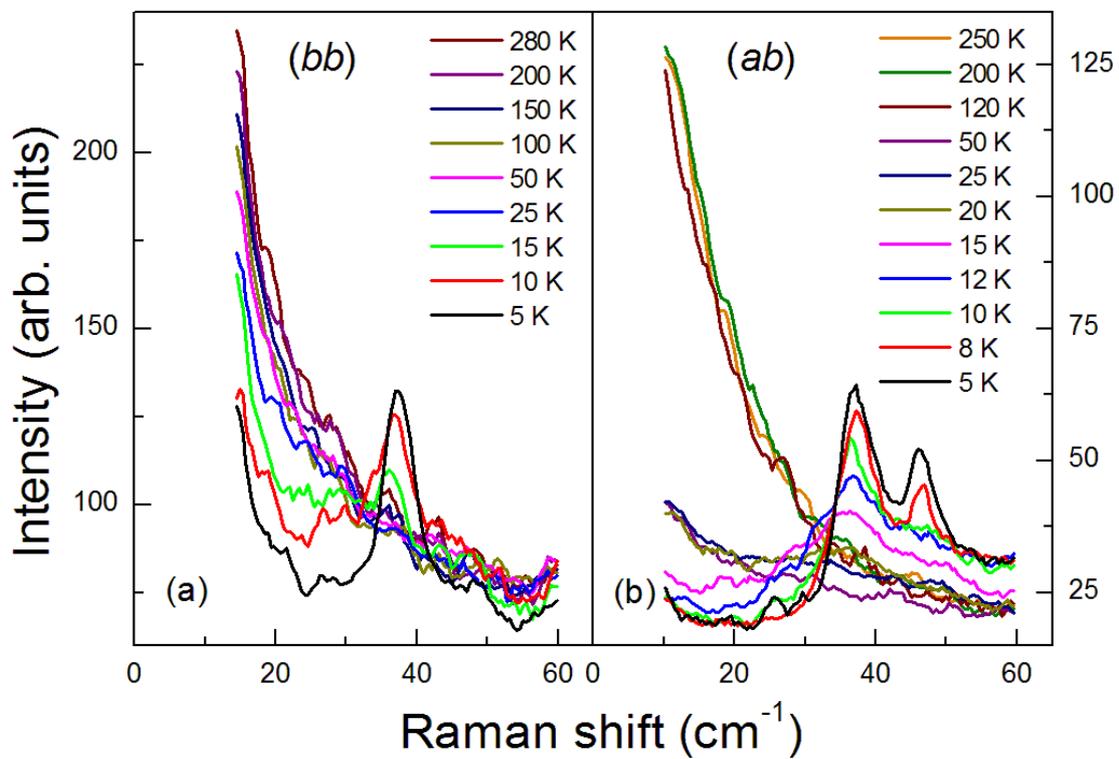

Figure 7. (Color online) Magnetic quasi-elastic scattering in α-TeVO$_4$ that evolves into finite energy modes for low temperatures and (bb) intrachain and (ab) crossed light polarization.

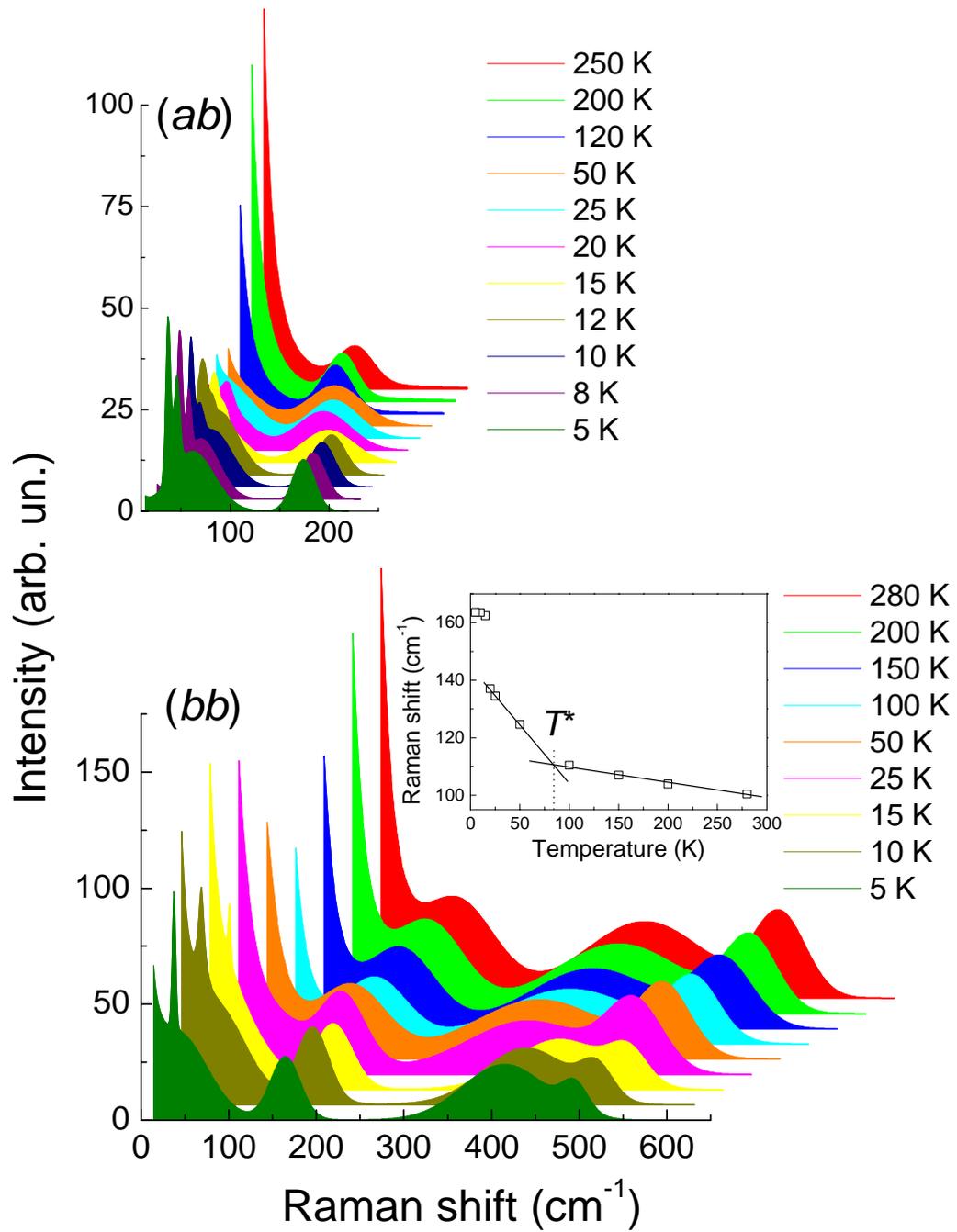

Figure 8. (Color online) Temperature variation of the magnetic Raman scattering in α-TeVO$_4$. The inset shows the temperature dependence of the frequency position of the band at ~100 cm$^{-1}$.

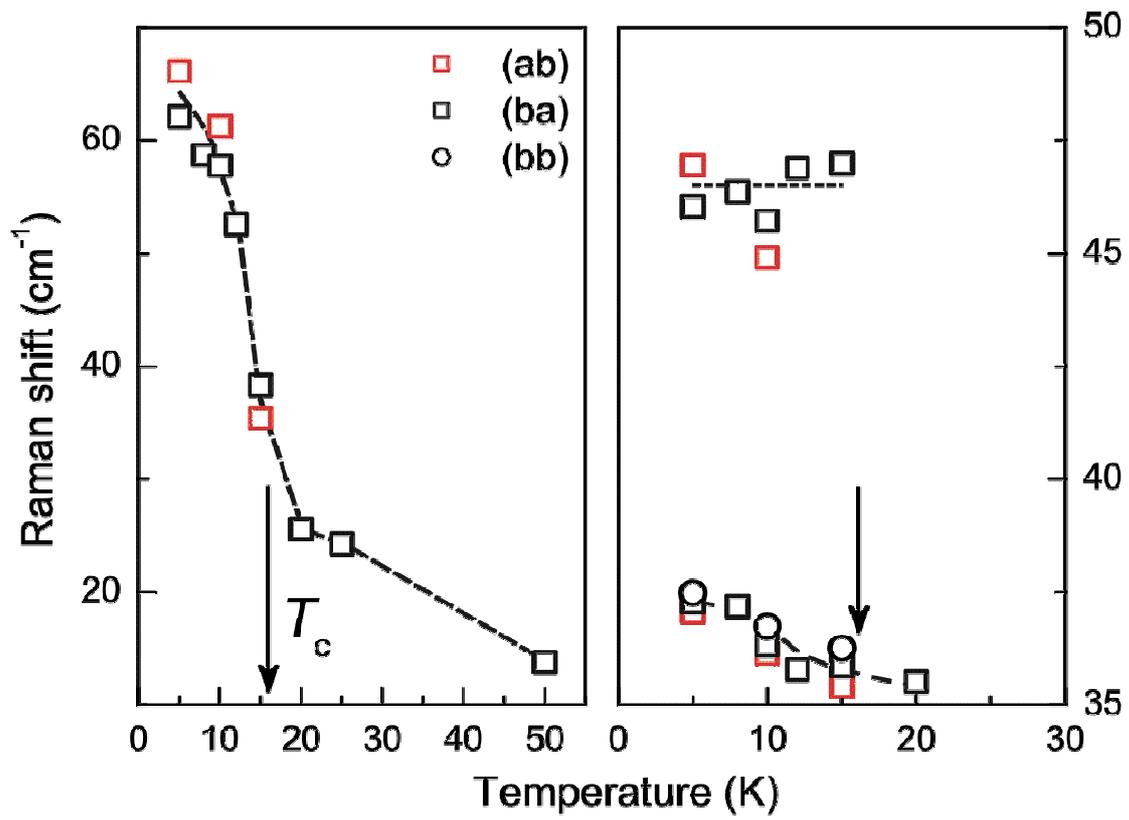

Figure 9. (Color online) Temperature dependence of the peak frequencies of the magnetic excitations in α-TeVO$_4$. Dashed lines are guides to the eye.